\documentclass[twocolumn,amsmath,amssymb]{revtex4-2}

\usepackage{graphicx}
\usepackage{hyperref}
\usepackage{braket}
\usepackage{bbm}
\usepackage[utf8]{inputenc}

\newcommand{\hankel}[1]{H^{(#1)}_n}

\usepackage[svgnames]{xcolor}
\usepackage{ulem}
\begin{document}

\title{Tractor beams with optimal pulling force using structured waves}

\author{Michael Horodynski}
\email{michael.horodynski@gmail.com}
\author{Tobias Reiter}
\author{Matthias Kühmayer}
\author{Stefan Rotter}
\affiliation{Institute for Theoretical Physics, Vienna University of Technology (TU Wien), A-1040 Vienna, Austria}

\begin{abstract}
    Moving objects with optical or acoustical waves is a topic both of fundamental interest and of importance for a range of practical applications. One particularly intriguing example is the tractor beam, which pulls an object toward the wave's source, in opposition to the wave's momentum. In this study, we introduce a protocol that enables the identification of wave states that produce the optimal tractor force for arbitrary objects. Our method relies solely on the solution of a simple eigenvalue problem involving the system's measurable scattering matrix. Using numerical simulations, we demonstrate the efficacy of this wavefront shaping protocol for a representative set of different targets. Moreover, we show that the diffractive nature of waves enables the possibility of a tractor beam, that works even for targets where a geometric optics approach fails to explain the pulling forces.
\end{abstract}

\maketitle

The widespread implementation of wave shaping tools in optics and acoustics has allowed researchers to create waves with a diverse set of interesting and often counter-intuitive properties \cite{mosk_controlling_2012,bliokh_roadmap_2023,rotter_light_2017,cao_shaping_2022,gigan_roadmap_2022}. These properties include propagation through complex media \cite{gerardin_full_2014,sarma_control_2016} and focusing inside and behind disordered materials \cite{horstmeyer_guidestar-assisted_2015,vellekoop_focusing_2007}, self-bending-airy beams \cite{efremidis_abruptly_2010}, optical tweezers \cite{ashkin_observation_1986,block_bead_1990,butaite_indirect_2019} and radiation pressure cooling \cite{magrini_real-time_2021,hupfl_optimal_2023}, to name just a few. Especially in the field of micromanipulation, the controlled movement of objects has been implemented with a remarkable level of efficiency. One particular example of this is the demonstration of a volumetric display using acoustic trapping \cite{hirayama_volumetric_2019}. Another is a study in which the authors demonstrate ultrasound beams that can levitate and steer solid objects in the urinary bladders of live pigs \cite{ghanem_noninvasive_2020}. A special class of wave-states that has received significant attention in this regard are the so-called tractor beams, which pull objects towards their source despite the wave's momentum being oriented in the opposite direction \cite{chen_optical_2011,novitsky_single_2011,brzobohaty_experimental_2013,li_optical_2020}.

These tractor beams have meanwhile been studied theoretically and experimentally both in the acoustic and optical domains. In the Rayleigh regime of scattering (target much smaller than the wavelength), studies were carried out that identify wave-states that pull particles towards their source by carefully balancing the intensity gradient and phase gradient force \cite{yevick_tractor_2016,abdelaziz_acoustokinetics_2020}. So far, however, tractor beams for extended objects have only been generated through numerical optimization of the wavefront \cite{marzo_holographic_2015}, heuristic design of the object and the wavefront \cite{demore_acoustic_2014,shvedov_long-range_2014,li_optical_2019}, and by exploiting chirality \cite{ding_realization_2014,fernandes_optical_2015}. The most difficult situation is when the object has Dirichlet boundary conditions since then waves perfectly bounce off it. So far this has been avoided by, e.g., including absorptive elements in the target \cite{demore_acoustic_2014}. Recently, evidence has emerged that not the design of the object but optimization of the incoming wavefront is the larger lever in light-matter interaction \cite{kuang_maximal_2020}. This is especially important since restricting tractor beams to only a small subset of engineered objects would be a considerable limitation. Thus, to fully unlock the power of tractor beams, an operational and practically implementable procedure is required to identify the optimal wavefront for pulling any given object to the wave source.

\begin{figure}[t!]
    \centering
    \includegraphics[width=0.5\textwidth]{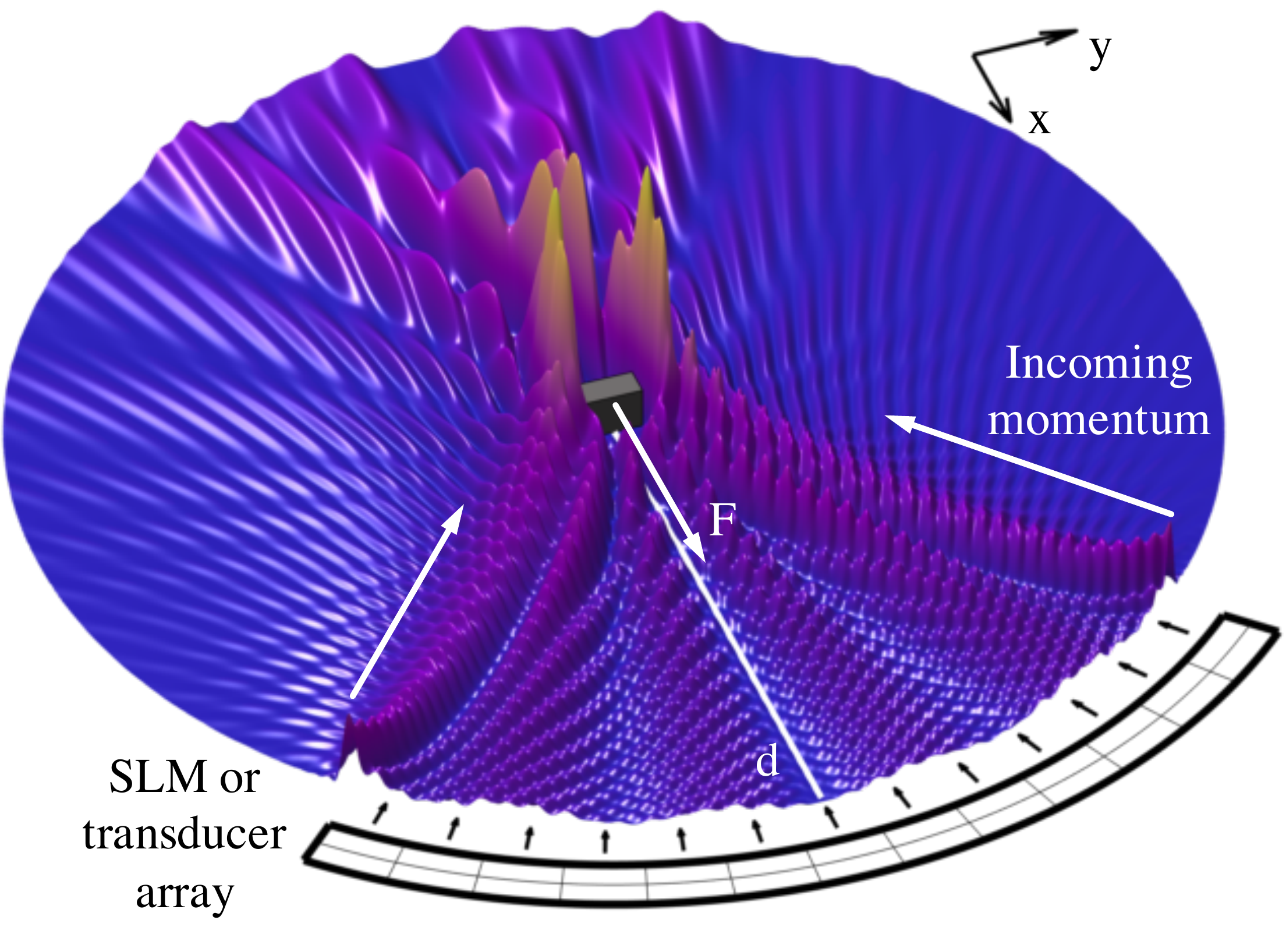}
    \caption{Illustration of the concept. In the circular scattering region of radius $R=20\lambda$, we place the object on which to exert pulling forces at a distance $d$ to the wave source. We restrict the corresponding region from which we generate waves to one-quarter of the circular boundary. Outgoing boundary conditions are implemented by a perfectly matched layer surrounding the scattering region (not pictured).}
    \label{fig:1}
\end{figure}

In this paper, we present such an approach that allows us to find tailor-made optimal wavefronts providing the best possible pulling force. Generally speaking, we compute the applied force on a target by considering an operator that maps the incoming wave field (typically given as a vector of modal amplitudes) to the applied force \cite{liu_optimal_2019,kuang_maximal_2020}. This approach has the advantage that the optimal wavefront for maximal pulling (or pushing) is simply the eigenvector of this operator associated with the largest (or smallest) eigenvalue of the operator. The operator we use here is the generalized Wigner-Smith (GWS) operator, which has been introduced for optimal focusing \cite{ambichl_focusing_2017} and has meanwhile been applied for micromanipulation \cite{horodynski_optimal_2020,butaite_photon-efficient_2023}, information retrieval \cite{bouchet_maximum_2021} and inverse design \cite{horodynski_anti-reflection_2022}. For the case at hand, where we are interested in moving an object towards the wave source (chosen here to coincide with the positive $x$-direction), this operator $Q_x$, evaluated in an appropriately chosen basis of in- and outgoing far-field modes, satisfies the following eigenvalue equation
\begin{align} \label{eq:gws}
    Q_x \ket{u} = \left(K^\mathrm{in}_x - S^\dagger K^\mathrm{out}_xS\right) \ket{u} = \theta_x \ket{u},
\end{align}
where $S$ is the system's scattering matrix (in our case describing the scattering off a target in free space) and $K^\mathrm{in}_x$ ($K^\mathrm{out}_x$) is the operator that allows for the computation of the incoming (outgoing) wavefront's momentum in $x$-direction (the direction in which we want to push/pull). It is then easy to see that $Q_x$ measures the difference between the incoming and outgoing momentum of the wave. Due to momentum conservation, this difference is then applied to the target \cite{ambichl_focusing_2017,strasser_direct_2021}. The advantage of using the GWS operator instead of other approaches like the optical eigenmode approach \cite{mazilu_optical_2011} is, that it only requires far-field information (the $S$-matrix), while the optical eigenmodes need a relation between far-field and near field in the target plane. We can understand Eq.~\eqref{eq:gws} not only as the difference between incoming and outgoing momenta as measured by $K_x^\mathrm{in}$ and $K_x^\mathrm{out}$, but also as an infinitesimal shift of the scattering matrix (since $Q_x=-\mathrm{i}S^\dagger \partial_x S$) \cite{ambichl_focusing_2017}. This has the consequence that $Q_x$ can not only be evaluated by shifting the target (to determine the derivative $\partial_x$) but also by shifting the spatial light modulator (SLM) or transducer array used to shape the incoming wave (akin to the equivalence between active and passive transformations).

To demonstrate our approach for constructing optimal tractor beam states, we first compute a unitary scattering matrix (for which the GWS operator is Hermitian, hence the eigenstate corresponding to the most positive eigenvalue then describes the input field of the optimal tractor beam). Since procedures for how to set up this scattering matrix in an appropriate basis (in free space) don't seem to be available in the literature, we present a comprehensive description of our solution in the following, with the details being available in Appendix~\ref{ap:a} and the code being published alongside this work \cite{horodynski_open_2023}. Restricting our analysis to two spatial dimensions, our starting point is the  scalar Helmholtz equation in polar coordinates, $[\Delta +k^2 \varepsilon(\vec{r})]\psi(\vec{r})=0$, in a circular region (see Fig.~\ref{fig:1}), which we solve numerically using an open-source finite-element method (NGSolve) \cite{schoberl_netgen_1997,schoberl_c11_2014}. Here $\Delta$ is the Laplacian in polar coordinates, $k$ is the wavenumber, $\varepsilon(\vec{r})$ is the spatially varying dielectric constant, $\psi(\vec{r})$ is the scalar wave and $\vec{r}=(\rho,\varphi)^T$ is the position vector. 

\begin{figure}[t!]
    \centering
    \includegraphics[width=\columnwidth]{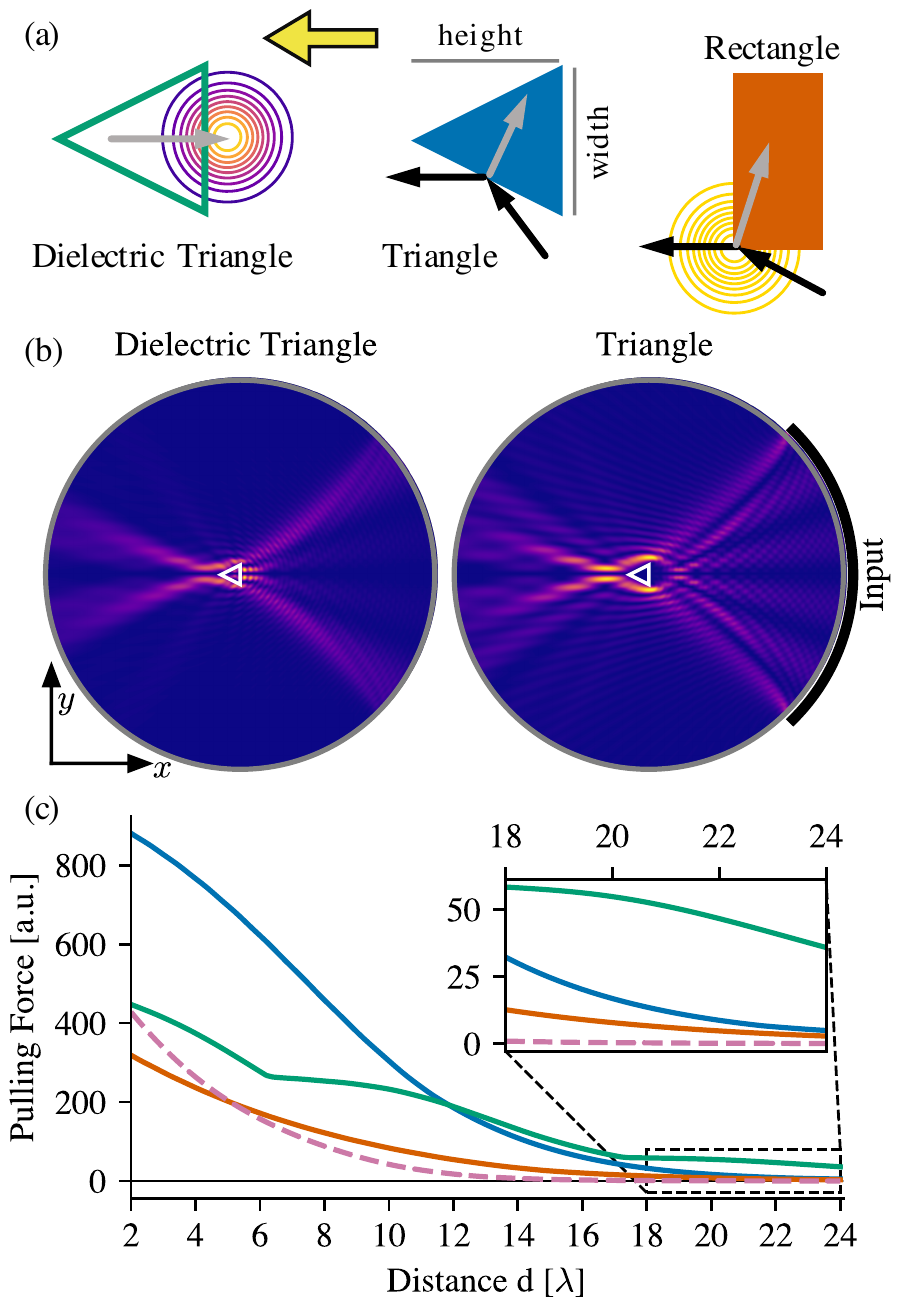}
    \caption{(a) Sketches of the targets used, which are a dielectric triangle (green) and 2 targets with Dirichlet boundary conditions: a triangle (blue) and a rectangle (orange). The black arrows mark the momentum of the incoming and scattered waves and the gray arrow is the resulting momentum transfer onto the target. Both the width of the triangles as well as their height measure $2\lambda$. The height of the rectangle is equal to $1\lambda$ and the width is $2\lambda$. The incoming waves impinge onto the targets from the right (yellow arrow). (b) Intensity distribution of the optimal tractor beam for the dielectric triangle (left) and the triangle with hard walls (right) at a distance of $d=20\lambda$ to the source. (c) The pulling force resulting from the optimal wave state for each distance over the distance from the source for the targets depicted in (a). Additionally, we also consider a rectangle with a width of $12\lambda$ (dashed violet line). The inset details the results at large distances from the source (marked by the dashed black line).}
    \label{fig:2}
\end{figure}

\begin{figure*}[t!]
    \centering
    \includegraphics[width=\textwidth]{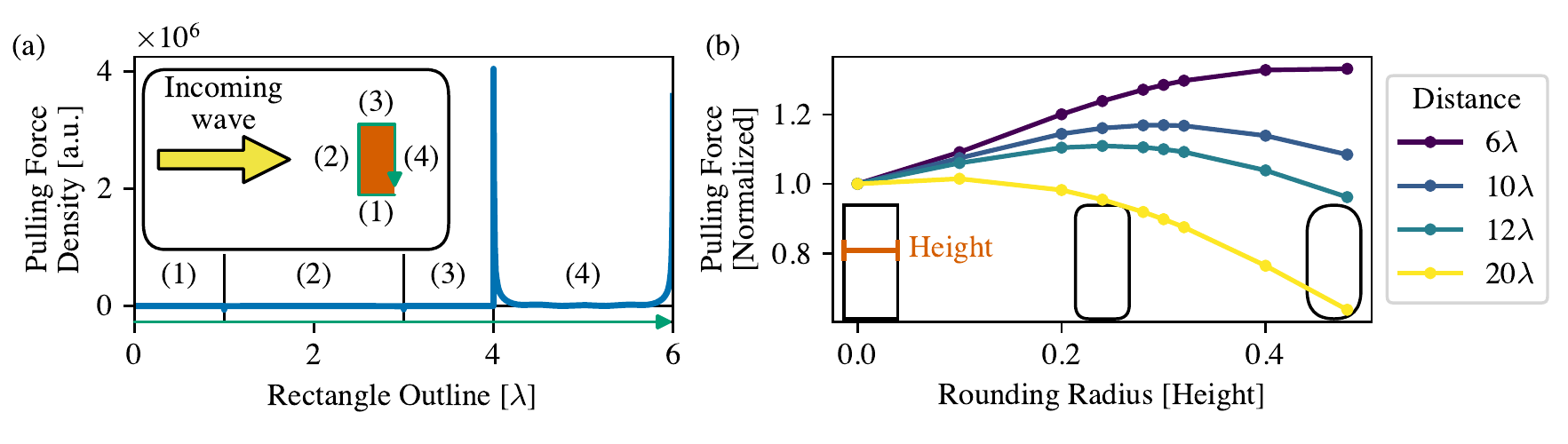}
    \caption{Diffraction forces on a rectangle. (a) Distribution of the pulling force around the rectangle with Dirichlet boundary conditions at a distance of $d=8\lambda$ to the source. (1) and (3) are the sides orthogonal to the pulling direction while (2) is facing towards and (4) is facing away from the source. The force, mainly located on the corners of side (4), is stronger there, resulting in a tractor beam. (b) Pulling force depending on the rounding radius for different distances from the source. The pulling force is normalized for each distance separately by its value at a rounding radius of zero.}
    \label{fig:3}
\end{figure*}

The fundamental solutions of the Helmholtz equation in the radial direction are Bessel functions of the first and second kind. From them, we can form by linear combination the Hankel functions of the first and second kind, representing outgoing and incoming waves in polar coordinates. The contribution of the angular variable $\varphi$ is given by $e^{i m \varphi}$, where $m$ numbers the mode. The unitary scattering matrix of the system is then computed by the following integral containing the numerical solution of the Helmholtz equation ($\psi_n$) resulting from a given input mode: $ S_{mn} = \int_0^{2\pi}  \frac{e^{\mathrm{i}m\varphi} \psi_n}{2\pi H^{(1)}_m(kR)} \mathrm{d} \varphi - \frac{\hankel{2}(kR)}{\hankel{1}(kR)}\delta_{m,-n} $. The fundamental solution of the Helmholtz equation also allows us to construct the incoming modes of this geometry as $\chi^\mathrm{in}_n = e^{\mathrm{i}n\varphi}\hankel{2}(k\rho)/(\sqrt{2\pi}|\hankel{2}(kR)|)$. Equipped with this basis we compute the elements of $K^\mathrm{in}_x$ by $[K^\mathrm{in}_x]_{mn} = -i\int_0^{2\pi} \mathrm{d}\varphi (\chi^\mathrm{in}_m)^* \partial_x\chi^\mathrm{in}_n$ and we note that $K^\mathrm{out}_x = K^\mathrm{in}_x$. 

In the rest of the paper, we fix the wavelength to $\lambda =R/20$, where $R$ is the radius of the scattering region. We also restrict the region from which we can send waves onto the target to a quarter of the scattering boundary (corresponding to a solid angle of $\pi/2$) so that all incoming waves have momentum directed opposite to the direction in which we want to pull the target. We do this by first transforming the GWS operator from the modal basis into the eigenbasis of the angle $\varphi$, which we get by solving the eigenvalue problem of the corresponding operator: $[\phi]_{mn} = \int_0^{2\pi} \mathrm{d}\varphi (\chi^\mathrm{in}_m)^* \varphi\chi^\mathrm{in}_n$. The eigenstates of $\phi$ are incoming waves that best approximate a point source at some location on the boundary, while also providing an orthogonal and complete basis. Expressing any state in this angular eigenbasis of this operator thus gives us the angular distribution of the state along the boundary of the scattering region. In this basis, we then only select contributions that lie within the allowed input aperture. We note here, that this only limits the angle in which we send waves into the system, but we still record all scattered waves. This procedure keeps the hermiticity of $Q_x$ such that we find the globally optimal tractor beam for a restricted input angle.

To show the power of the presented tool, we consider a representative set of targets: First, a dielectric triangle (refractive index $n=1.44$), where our calculations show that a focus on the target's front will execute an efficient pulling force by drawing the target to regions with higher intensity (see Fig.~\ref{fig:2}a for an illustration of the concept and Fig.~\ref{fig:2}b for the intensity distribution of the optimal tractor beam). Second, the already more challenging case of a triangle with fully reflecting boundary conditions (``hard walls'') is considered. Here, an intuitive ray optics picture suggests that optical pulling forces can be implemented by rays bouncing off the slanted sides of the triangle, such that their redirection results in a momentum transfer that pulls the target to the rays' source \cite{sukhov_negative_2011,demore_acoustic_2014} (see Fig.~\ref{fig:2}a for an illustration of the concept and Fig.~\ref{fig:2}b for the intensity distribution of the optimal tractor beam, which shows an appropriate redirection of the beams). Third, we consider the case of rectangles with hard walls of varying widths which presents a counter-intuitive scenario in the sense that optical pulling forces cannot be understood through ray optics. This is because all rays that are reflected from those sides of the rectangle, which are in a line of sight with the SLM or transducer array, can only result in pushing the rectangle away or in causing lateral displacement. Nevertheless, leveraging the full interferometric nature, e.g., of electromagnetic or acoustic waves, our approach still finds the incoming state that exerts the maximal pulling force by an appropriate redirection of the incoming wave through diffraction at the rectangle's corners (see Fig.~\ref{fig:2}a for the illustration and Fig.~\ref{fig:1} for the intensity distribution which showcases the redirection of the wave). 

\begin{figure*}[t!]
    \centering
    \includegraphics[width=\textwidth]{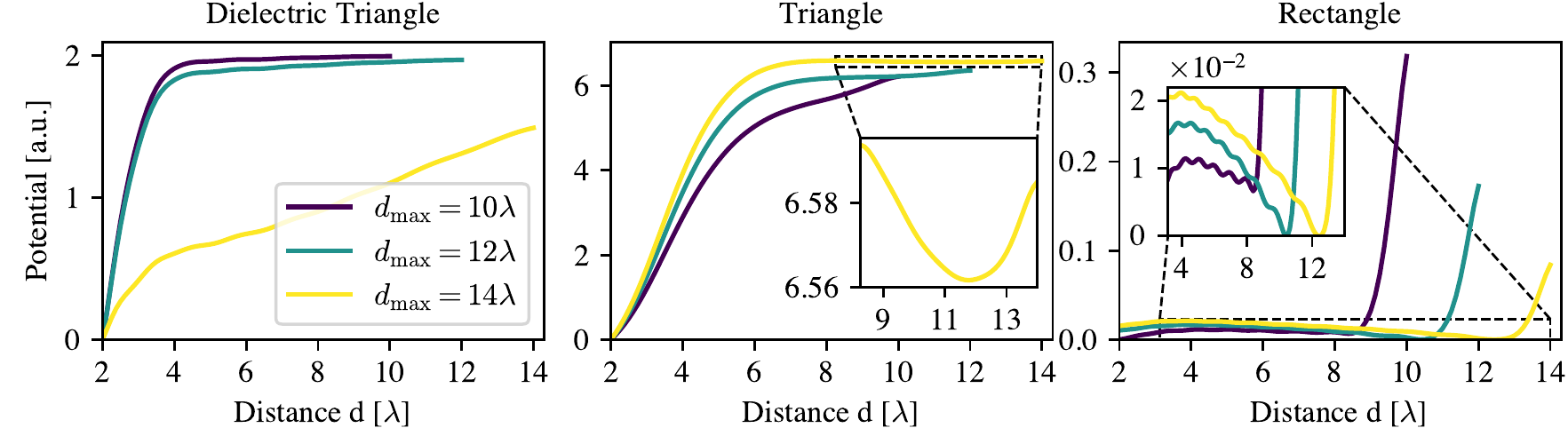}
    \caption{Performance of a static tractor beam for different maximal distances of the object (color scale). Depicted is the potential as a function of the distance $d$. In all cases, $d_\mathrm{min}=2\lambda$ and the distance between two evaluation points of $Q_x$ is $\lambda/2$, i.e. $d_{i+1}-d_i=\lambda/2$. The insets enlarge the area marked by the dashed rectangle and show local minima of the potential in which targets could get stuck.}
    \label{fig:4}
\end{figure*}

To elucidate this behavior, we first look at the force density over the boundary of the rectangle. Fig.~\ref{fig:3}a clearly shows that the majority of the force is applied very close to the rectangle's corners, suggesting that diffraction at these corners is responsible for the emergence of optical pulling forces. To further corroborate this statement, we show in Fig.~\ref{fig:3}b the strength of the optical pulling force as a function of how much we round the sharp corners of a rectangle. We discover that at short distances to the source, rectangles with a large rounding radius allow for a stronger pulling force than rectangles with sharp corners, while for long distances this effect is reversed, i.e., sharp corners allow the largest pulling forces. We attribute this behavior to the ease with which waves can focus at short distances at the corners' backside allowing us to employ the intuitive picture of specularly reflected rays at the corners on the distal end of the rectangle, instead of a more complicated wave-based explanation relying on diffraction. For longer distances, such a focus is harder to achieve and in this case, the pulling force is induced by wave diffraction -- which is more pronounced at sharp corners of the rectangle's corners.

In Fig.~\ref{fig:2}c we show the maximal pulling force for all of the individual targets, evaluated at different distances to the origin. The depicted simulation results confirm our intuition from above in the sense that the pulling force for the dielectric target decays the slowest with increasing distance. For the targets with hard walls, the slowest decay is observed for the triangle, while the wider rectangle shows the fastest decay – demonstrating at the same time that a narrowing numerical aperture is detrimental to the strength of optical pulling forces.

So far we have discussed tractor beams optimized for each distance of the object to the source individually. Thus, to pull a target closer to the source, the incoming wave has to continuously change. This is akin to optical conveyor beams, which can not only trap an object but also pull them in by changing the phase of the beam \cite{ruffner_optical_2012}. In the following, we also demonstrate that wavefront shaping allows for the creation of a ``static'' tractor beam, which exerts a pulling force on the target irrespective of its current distance from the source. This static tractor beam only requires a single unchanging pattern on the SLM. We find this wavefront by considering the eigenstates of the following operator
\begin{equation}
    Q_{x}^\mathrm{cont} = \sum_{d_\mathrm{min}}^{d_\mathrm{max}} \frac{1}{\theta_\mathrm{max}\left(d_i\right)} Q_{x}\left(d_i\right), \label{eq:ctb}
\end{equation}
which is the sum of all GWS operators, spaced by some distance along the way, weighted with their maximal eigenvalues (i.e., the one that indicates the maximal strength of the pulling force). By $d_\mathrm{min}$ ($d_\mathrm{max}$) we denote the minimal (maximal) distance to the source we consider. We also note here that the distance between the points in space at which two adjacent GWS operators are evaluated is evenly spaced. The weighted sum is used because otherwise, the GWS operators at short distances would dominate since at short distances the applied pushing and pulling forces are much stronger (see Fig.~\ref{fig:2}c). In Fig.~\ref{fig:4} we show the performance of the static tractor beam found by solving the eigenvalue problem of $Q_{x}^\mathrm{cont}$. We see that with this protocol it is indeed possible to construct a static tractor beam for different targets (such as for the triangle and rectangle with hard walls as well as for the dielectric triangle). We note that the potential created by the static tractor beam can feature shallow local minima, in which that target could get trapped. However, in the case of the rectangle, if it starts at an appropriate distance it can garner enough energy to just roll over such a potential well.

We also compare different $d_\mathrm{max}$ to investigate how the performance of a static tractor beam is affected by the maximal distance we demand from it. Our simulations uncover that depending on which target we consider, there is a different maximal distance for which our protocol finds a static tractor beam. The best performance is found when considering the dielectric triangle and the worst is when considering the triangle with hard walls. This result is surprising since the triangle with hard walls exhibits a greater range at which it is possible to engineer a tractor beam state compared to the rectangle (see Fig.~\ref{fig:2}c). 

We note here that Eq.~\eqref{eq:ctb} is not the only choice to find a static tractor beam and we, therefore, tried two alternatives, which, however, did not surpass the presented method in their performance: The first is to replace the division by the extremal eigenvalue with a division by the trace of $Q_x$, which would then give the optimal state for this particular weighted sum. The second is formulating a (constrained) numerical optimization problem and solving it with an appropriate numerical library. This gives us greater freedom in fixing the properties of the wave state at the prize that the optimization problem will be non-convex (because $Q_x$ is not positive definite) and thus the solution will not be guaranteed to converge to the global optimum \cite{boyd_convex_2004}.

To conclude, we demonstrate a protocol that finds for arbitrary objects a wavefront that exerts the optimal pulling force. We also uncover the mechanisms responsible for the transfer of the tractor force and show in which way they depend on the shape of the target. Furthermore, we propose a scheme for a static tractor beam, i.e., a wave state resulting from an unchanging SLM pattern that exerts a pulling force onto the target, irrespective of the target's distance from the wave source. An interesting open question building on the insights presented here is the concurrent optimization of the wavefront and of the object's shape. Our approach would be ideally suited to investigate this since it provides both the optimal wavefront as well as the gradient of the cost function with respect to changes in the geometry \cite{horodynski_anti-reflection_2022}.

The computational results were achieved using the Vienna Scientific Cluster (VSC).

\appendix
\section{Details of the numerical implementation}\label{ap:a}

\begin{figure*}[t!]
    \centering
    \includegraphics[width=\textwidth]{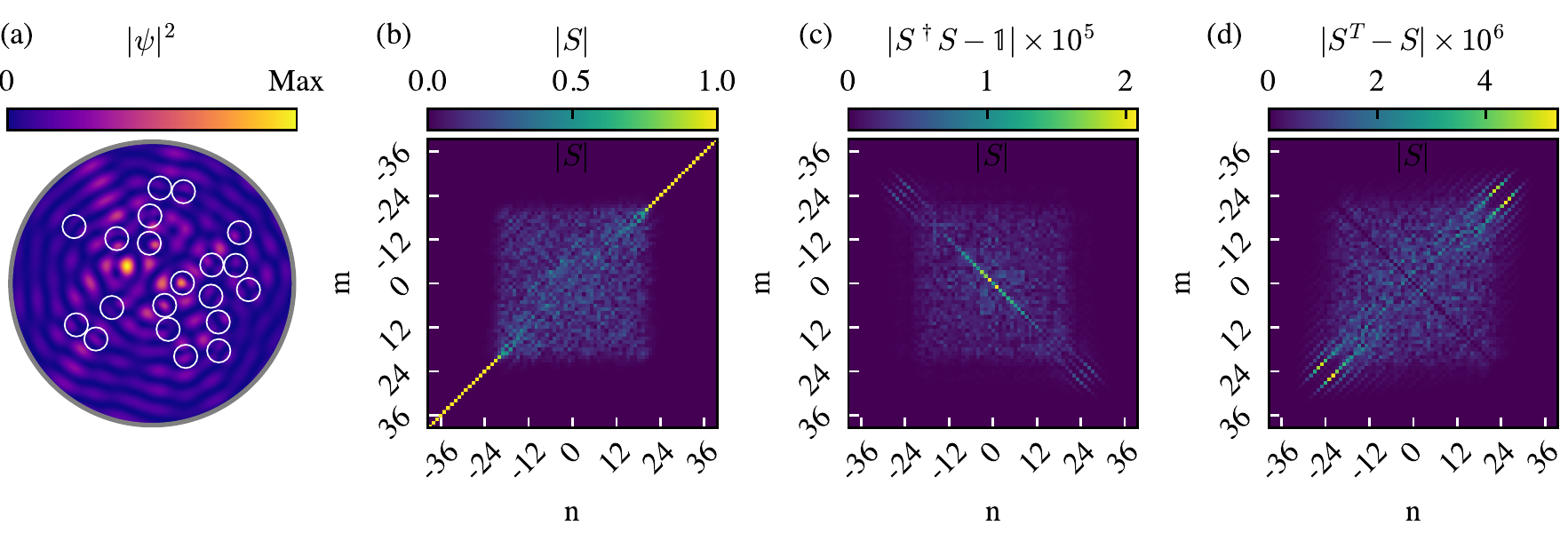}
    \caption{Properties of the $S$ matrix. (a) Spatial intensity distribution of an incoming field in the fundamental mode ($n=0$) in a geometry consisting of 20 randomly placed Teflon scatterers ($\varepsilon=2.0736$) with radius $0.33\lambda$. The radius of the scattering region is $R=4\lambda$, which corresponds to 39 modes. (b) Absolute value of the scattering matrix entries for the system depicted in (a). (c) Absolute value of the deviation from unitarity ($S^\dagger S - \mathbbm{1}$). (d) Absolute value of the deviation from transposition symmetry ($S^T-S$).}
    \label{fig:a1}
\end{figure*}

Here, we show how to construct a unitary scattering matrix for an open system with a circular boundary in the framework of the finite-element tool NGSolve \cite{schoberl_netgen_1997,schoberl_c11_2014}. Our starting point is the most general solution of the Helmholtz equation in polar coordinates containing propagating waves:
\begin{align}
    \psi = \sum_{n\in\mathbb{Z}} \left[ \alpha_n H_n^{(2)}(k\rho) + \beta_n H_n^{(1)}(k\rho) \right] \left(\gamma_n e^{\mathrm{i}n \varphi} + \delta_n e^{-\mathrm{i}n \varphi} \right),
\end{align}
where outgoing and incoming waves are represented by the Hankel functions of the first and second kind ($H^{(1)}_n$ and $H^{(2)}_n$), respectively, since 
\begin{align} \label{eq:flux}
    \Re \left[ -\mathrm{i} {\hankel{1,2}}^*(k\rho) \partial_\rho \hankel{1,2}(k\rho) \right] & = \pm\frac{2}{\pi \rho} \;\forall n, \;\forall k\rho,
\end{align}
where $\Re$ denotes taking the real part and we note here that the radial flux is independent of $n$. This immediately gives us a complete and orthonormal basis on which we can construct the scattering matrix. The only remaining task is then to get every constant pre-factor in the numerical implementation and computation of the $S$-matrix exactly right in order to have a flux-conserving and thus unitary $S$-matrix, as well as an $S$-matrix that respects transposition symmetry. 

The first aspect of this task is to carefully look at the source term ($f$, which we have omitted so far) in the Helmholtz equation: 
\begin{equation}\label{eq:pde_s}
    [\Delta +k^2 \varepsilon(\vec{r})]\psi(\vec{r})=-f(\vec{r}).
\end{equation}
In the concrete implementation of this work $f$ is located along the circular boundary of the scattering region, i.e., $f(\rho,\varphi)=\delta(\rho-R)h(\varphi)=\delta(\rho-R)\sum_n c_n e^{\mathrm{i}n\varphi}$, where $c_n$ are the entries of a vector of modal amplitudes that describes the incoming wave. We now want to fix $c_n$ to get excitations of the form
\begin{align}
    \psi_I = & \sum_n a_n \hankel{2}(k\rho) e^{\mathrm{i}n\varphi} \;\mathrm{for}\; \rho < R, \label{eq:psiI} \\
    \psi_O = & \sum_n b_n \hankel{1}(k\rho) e^{\mathrm{i}n\varphi} \;\mathrm{for}\; \rho > R,
\end{align}
where $\psi_I$ and $\psi_O$ represent waves inside and outside the circular boundary that both propagate away from the source. We choose this particular form of $\psi_I$ to have an isotropic source when considering $|\psi_I|$. By demanding that $\psi_I(R)=\psi_O(R)$ (continuity of the solution) we find that $a_n\hankel{2}(kR)=b_n\hankel{1}(kR)$. To connect $a_n$ and $b_n$ to $c_n$ we then plug our particular form of the source into the (vacuum) Helmholtz equation multiply it by $e^{-\mathrm{i}m\varphi}$ and integrate from $R-\epsilon$ to $R+\epsilon$ ($\epsilon$ being a small number, which we later take to zero) and from $0$ to $2\pi$. This results in $c_n =- 4\mathrm{i}a_n/[\pi R H^{(1)}_n(kR)]$, which is central to the computation of a unitary scattering matrix, since for a numerical solution of Eq.~\eqref{eq:pde_s} we get excitations of the form~\eqref{eq:psiI}. There is also a second consequence of the formula connecting the $a_n$'s and $c_n$'s: It imposes a (heuristic) cutoff for the number of modes, since Hankel functions of constant $kR$ but increasing $n$ increase in absolute value, such that for some $n$, $c_n$ is effectively zero. 

Equipped with the necessary knowledge of the source's exact form, we can obtain a unitary $S$-matrix from a numerical solution of the Helmholtz equation ($\psi_n$) for an incoming cylindrical wave with amplitude one in mode $n$:
\begin{equation}\label{eq:S}
    S_{mn} =  \int_0^{2\pi} \frac{e^{\mathrm{i}m\varphi} \psi_n}{2\pi H^{(1)}_m(kR)} \mathrm{d} \varphi - \frac{\hankel{2}(kR)}{\hankel{1}(kR)}\delta_{m,-n} .
\end{equation}
In the above formula, we project $\psi_n$ onto $e^{\mathrm{i}m\varphi}$, since the outgoing channels are time-reversed incoming channels, i.e. the complex conjugate of the incoming mode \cite{liu_optimal_2019}. In order to correctly project onto the outgoing channels ($e^{-\mathrm{i}m\varphi}$) we then need to complex conjugate a second time. The factor in front of the integral in Eq.~\eqref{eq:S} is placed to ensure the correct normalization in both phase and amplitude for each element of $S$. It is, however, not a flux normalization factor, since every mode we consider carries the same flux [see Eq.~\eqref{eq:flux}]. In other words, the weighing of different fluxes for different modes is not necessary. To avoid including the incoming radiation in the computation of $S$, we must also subtract from each element of the anti-diagonal (which are the elements of $S_{mn}$ for which $m=-n$) the term $H^{(2)}_n(kR)/H^{(1)}_m(kR)$. 

In Fig.~\ref{fig:a1} we plot a system of randomly placed scatterers and its associated scattering matrix. We see that $S$ is unitary (since there is neither loss nor gain present) and transposition symmetric up to the numerical error. These are the fundamental symmetries of a scattering matrix \cite{rotter_light_2017}, indicating that our computation is correct. Fig.~\ref{fig:a1}b also shows that for some incoming modes, the non-zero elements of $S$ are located on the anti-diagonal. We also note that when the system is empty, all non-zero elements of $S$ are located on the anti-diagonal (not shown). This can be attributed to the fact that in an empty system, the solution is given by $\psi_n=[\hankel{1}(k\rho)+\hankel{2}(k\rho)]e^{\mathrm{i}n\varphi}=2J_n(k\rho)e^{\mathrm{i}n\varphi}$, which cannot be expressed as the sum of the incoming and outgoing modes, as doing so would violate the conservation of angular momentum.

Lastly, we also discuss the form of the basis state from which we can compute any operator, like $\phi$ and $K_x$. Since the incoming waves are represented by $e^{\mathrm{i}n\varphi} \hankel{2}(k \rho)$, the incoming modes are given by
\begin{equation}
    \chi_n^\mathrm{in} = \frac{e^{\mathrm{i}n\varphi} \hankel{2}(k \rho)}{\sqrt{2\pi} |\hankel{2}(kR)|},
\end{equation}
where the terms in the denominator are necessary to have orthonormal modes at $\rho=R$.

After computing the elements of $K_x^\mathrm{in}$ using $\chi_n^\mathrm{in}$ and in turn constructing the GWS operator with them, there is also secondary use for them: When we have computed the $S$-matrix for an arbitrary object at an arbitrary position $x$ within the system we can calculate the scattering matrix at a position shifted by $\Delta x$ by considering \cite{ambichl_focusing_2017}
\begin{equation}
    S(x + \Delta x) = e^{-\mathrm{i}K^\mathrm{in}_x \Delta x} S(x) e^{\mathrm{i}K^\mathrm{in}_x \Delta x} .
\end{equation}

\bibliographystyle{apsrev4-2}

\begin{thebibliography}{44}%
\makeatletter
\providecommand \@ifxundefined [1]{%
 \@ifx{#1\undefined}
}%
\providecommand \@ifnum [1]{%
 \ifnum #1\expandafter \@firstoftwo
 \else \expandafter \@secondoftwo
 \fi
}%
\providecommand \@ifx [1]{%
 \ifx #1\expandafter \@firstoftwo
 \else \expandafter \@secondoftwo
 \fi
}%
\providecommand \natexlab [1]{#1}%
\providecommand \enquote  [1]{``#1''}%
\providecommand \bibnamefont  [1]{#1}%
\providecommand \bibfnamefont [1]{#1}%
\providecommand \citenamefont [1]{#1}%
\providecommand \href@noop [0]{\@secondoftwo}%
\providecommand \href [0]{\begingroup \@sanitize@url \@href}%
\providecommand \@href[1]{\@@startlink{#1}\@@href}%
\providecommand \@@href[1]{\endgroup#1\@@endlink}%
\providecommand \@sanitize@url [0]{\catcode `\\12\catcode `\$12\catcode
  `\&12\catcode `\#12\catcode `\^12\catcode `\_12\catcode `\%12\relax}%
\providecommand \@@startlink[1]{}%
\providecommand \@@endlink[0]{}%
\providecommand \url  [0]{\begingroup\@sanitize@url \@url }%
\providecommand \@url [1]{\endgroup\@href {#1}{\urlprefix }}%
\providecommand \urlprefix  [0]{URL }%
\providecommand \Eprint [0]{\href }%
\providecommand \doibase [0]{https://doi.org/}%
\providecommand \selectlanguage [0]{\@gobble}%
\providecommand \bibinfo  [0]{\@secondoftwo}%
\providecommand \bibfield  [0]{\@secondoftwo}%
\providecommand \translation [1]{[#1]}%
\providecommand \BibitemOpen [0]{}%
\providecommand \bibitemStop [0]{}%
\providecommand \bibitemNoStop [0]{.\EOS\space}%
\providecommand \EOS [0]{\spacefactor3000\relax}%
\providecommand \BibitemShut  [1]{\csname bibitem#1\endcsname}%
\let\auto@bib@innerbib\@empty
\bibitem [{\citenamefont {Mosk}\ \emph {et~al.}(2012)\citenamefont {Mosk},
  \citenamefont {Lagendijk}, \citenamefont {Lerosey},\ and\ \citenamefont
  {Fink}}]{mosk_controlling_2012}%
  \BibitemOpen
  \bibfield  {author} {\bibinfo {author} {\bibfnamefont {A.~P.}\ \bibnamefont
  {Mosk}}, \bibinfo {author} {\bibfnamefont {A.}~\bibnamefont {Lagendijk}},
  \bibinfo {author} {\bibfnamefont {G.}~\bibnamefont {Lerosey}},\ and\ \bibinfo
  {author} {\bibfnamefont {M.}~\bibnamefont {Fink}},\ }\href
  {https://doi.org/10.1038/nphoton.2012.88} {\bibfield  {journal} {\bibinfo
  {journal} {Nature Photonics}\ }\textbf {\bibinfo {volume} {6}},\ \bibinfo
  {pages} {283} (\bibinfo {year} {2012})}\BibitemShut {NoStop}%
\bibitem [{\citenamefont {Bliokh}\ \emph {et~al.}(2023)\citenamefont {Bliokh},
  \citenamefont {Karimi}, \citenamefont {Padgett}, \citenamefont {Alonso},
  \citenamefont {Dennis}, \citenamefont {Dudley}, \citenamefont {Forbes},
  \citenamefont {Zahedpour}, \citenamefont {Hancock}, \citenamefont
  {Milchberg}, \citenamefont {Rotter}, \citenamefont {Nori}, \citenamefont
  {Özdemir}, \citenamefont {Bender}, \citenamefont {Cao}, \citenamefont
  {Corkum}, \citenamefont {Hernandez-Garcia}, \citenamefont {Ren},
  \citenamefont {Kivshar}, \citenamefont {Silveirinha}, \citenamefont
  {Engheta}, \citenamefont {Rauschenbeutel}, \citenamefont {Schneeweiss},
  \citenamefont {Volz}, \citenamefont {Leykam}, \citenamefont {Smirnova},
  \citenamefont {Rong}, \citenamefont {Wang}, \citenamefont {Hasman},
  \citenamefont {Picardi}, \citenamefont {Zayats}, \citenamefont
  {Rodriguez-Fortuno}, \citenamefont {Yang}, \citenamefont {Ren}, \citenamefont
  {Khanikaev}, \citenamefont {Alù}, \citenamefont {Brasselet}, \citenamefont
  {Shats}, \citenamefont {Verbeeck}, \citenamefont {Schattschneider},
  \citenamefont {Sarenac}, \citenamefont {Cory}, \citenamefont {Pushin},
  \citenamefont {Birk}, \citenamefont {Gorlach}, \citenamefont {Kaminer},
  \citenamefont {Cardano}, \citenamefont {Marrucci}, \citenamefont {Krenn},\
  and\ \citenamefont {Marquardt}}]{bliokh_roadmap_2023}%
  \BibitemOpen
  \bibfield  {author} {\bibinfo {author} {\bibfnamefont {K.~Y.}\ \bibnamefont
  {Bliokh}}, \bibinfo {author} {\bibfnamefont {E.}~\bibnamefont {Karimi}},
  \bibinfo {author} {\bibfnamefont {M.~J.}\ \bibnamefont {Padgett}}, \bibinfo
  {author} {\bibfnamefont {M.~A.}\ \bibnamefont {Alonso}}, \bibinfo {author}
  {\bibfnamefont {M.~R.}\ \bibnamefont {Dennis}}, \bibinfo {author}
  {\bibfnamefont {A.}~\bibnamefont {Dudley}}, \bibinfo {author} {\bibfnamefont
  {A.}~\bibnamefont {Forbes}}, \bibinfo {author} {\bibfnamefont
  {S.}~\bibnamefont {Zahedpour}}, \bibinfo {author} {\bibfnamefont {S.~W.}\
  \bibnamefont {Hancock}}, \bibinfo {author} {\bibfnamefont {H.~M.}\
  \bibnamefont {Milchberg}}, \bibinfo {author} {\bibfnamefont {S.}~\bibnamefont
  {Rotter}}, \bibinfo {author} {\bibfnamefont {F.}~\bibnamefont {Nori}},
  \bibinfo {author} {\bibfnamefont {S.~K.}\ \bibnamefont {Özdemir}}, \bibinfo
  {author} {\bibfnamefont {N.}~\bibnamefont {Bender}}, \bibinfo {author}
  {\bibfnamefont {H.}~\bibnamefont {Cao}}, \bibinfo {author} {\bibfnamefont
  {P.~B.}\ \bibnamefont {Corkum}}, \bibinfo {author} {\bibfnamefont
  {C.}~\bibnamefont {Hernandez-Garcia}}, \bibinfo {author} {\bibfnamefont
  {H.}~\bibnamefont {Ren}}, \bibinfo {author} {\bibfnamefont {Y.}~\bibnamefont
  {Kivshar}}, \bibinfo {author} {\bibfnamefont {M.~G.}\ \bibnamefont
  {Silveirinha}}, \bibinfo {author} {\bibfnamefont {N.}~\bibnamefont
  {Engheta}}, \bibinfo {author} {\bibfnamefont {A.}~\bibnamefont
  {Rauschenbeutel}}, \bibinfo {author} {\bibfnamefont {P.}~\bibnamefont
  {Schneeweiss}}, \bibinfo {author} {\bibfnamefont {J.}~\bibnamefont {Volz}},
  \bibinfo {author} {\bibfnamefont {D.}~\bibnamefont {Leykam}}, \bibinfo
  {author} {\bibfnamefont {D.~A.}\ \bibnamefont {Smirnova}}, \bibinfo {author}
  {\bibfnamefont {K.}~\bibnamefont {Rong}}, \bibinfo {author} {\bibfnamefont
  {B.}~\bibnamefont {Wang}}, \bibinfo {author} {\bibfnamefont {E.}~\bibnamefont
  {Hasman}}, \bibinfo {author} {\bibfnamefont {M.~F.}\ \bibnamefont {Picardi}},
  \bibinfo {author} {\bibfnamefont {A.~V.}\ \bibnamefont {Zayats}}, \bibinfo
  {author} {\bibfnamefont {F.~J.}\ \bibnamefont {Rodriguez-Fortuno}}, \bibinfo
  {author} {\bibfnamefont {C.}~\bibnamefont {Yang}}, \bibinfo {author}
  {\bibfnamefont {J.}~\bibnamefont {Ren}}, \bibinfo {author} {\bibfnamefont
  {A.~B.}\ \bibnamefont {Khanikaev}}, \bibinfo {author} {\bibfnamefont
  {A.}~\bibnamefont {Alù}}, \bibinfo {author} {\bibfnamefont {E.}~\bibnamefont
  {Brasselet}}, \bibinfo {author} {\bibfnamefont {M.}~\bibnamefont {Shats}},
  \bibinfo {author} {\bibfnamefont {J.}~\bibnamefont {Verbeeck}}, \bibinfo
  {author} {\bibfnamefont {P.}~\bibnamefont {Schattschneider}}, \bibinfo
  {author} {\bibfnamefont {D.}~\bibnamefont {Sarenac}}, \bibinfo {author}
  {\bibfnamefont {D.~G.}\ \bibnamefont {Cory}}, \bibinfo {author}
  {\bibfnamefont {D.}~\bibnamefont {Pushin}}, \bibinfo {author} {\bibfnamefont
  {M.}~\bibnamefont {Birk}}, \bibinfo {author} {\bibfnamefont {A.}~\bibnamefont
  {Gorlach}}, \bibinfo {author} {\bibfnamefont {I.}~\bibnamefont {Kaminer}},
  \bibinfo {author} {\bibfnamefont {F.}~\bibnamefont {Cardano}}, \bibinfo
  {author} {\bibfnamefont {L.}~\bibnamefont {Marrucci}}, \bibinfo {author}
  {\bibfnamefont {M.}~\bibnamefont {Krenn}},\ and\ \bibinfo {author}
  {\bibfnamefont {F.}~\bibnamefont {Marquardt}},\ }\href
  {http://arxiv.org/abs/2301.05349} {\bibinfo {title} {Roadmap on structured
  waves}} (\bibinfo {year} {2023}),\ \bibinfo {note} {arXiv:2301.05349
  [physics, physics:quant-ph]}\BibitemShut {NoStop}%
\bibitem [{\citenamefont {Rotter}\ and\ \citenamefont
  {Gigan}(2017)}]{rotter_light_2017}%
  \BibitemOpen
  \bibfield  {author} {\bibinfo {author} {\bibfnamefont {S.}~\bibnamefont
  {Rotter}}\ and\ \bibinfo {author} {\bibfnamefont {S.}~\bibnamefont {Gigan}},\
  }\href {https://doi.org/10.1103/RevModPhys.89.015005} {\bibfield  {journal}
  {\bibinfo  {journal} {Reviews of Modern Physics}\ }\textbf {\bibinfo {volume}
  {89}},\ \bibinfo {pages} {015005} (\bibinfo {year} {2017})}\BibitemShut
  {NoStop}%
\bibitem [{\citenamefont {Cao}\ \emph {et~al.}(2022)\citenamefont {Cao},
  \citenamefont {Mosk},\ and\ \citenamefont {Rotter}}]{cao_shaping_2022}%
  \BibitemOpen
  \bibfield  {author} {\bibinfo {author} {\bibfnamefont {H.}~\bibnamefont
  {Cao}}, \bibinfo {author} {\bibfnamefont {A.~P.}\ \bibnamefont {Mosk}},\ and\
  \bibinfo {author} {\bibfnamefont {S.}~\bibnamefont {Rotter}},\ }\href
  {https://doi.org/10.1038/s41567-022-01677-x} {\bibfield  {journal} {\bibinfo
  {journal} {Nature Physics}\ }\textbf {\bibinfo {volume} {18}},\ \bibinfo
  {pages} {994} (\bibinfo {year} {2022})}\BibitemShut {NoStop}%
\bibitem [{\citenamefont {Gigan}\ \emph {et~al.}(2022)\citenamefont {Gigan},
  \citenamefont {Katz}, \citenamefont {de~Aguiar}, \citenamefont {Andresen},
  \citenamefont {Aubry}, \citenamefont {Bertolotti}, \citenamefont {Bossy},
  \citenamefont {Bouchet}, \citenamefont {Brake}, \citenamefont {Brasselet},
  \citenamefont {Bromberg}, \citenamefont {Cao}, \citenamefont {Chaigne},
  \citenamefont {Cheng}, \citenamefont {Choi}, \citenamefont {Čižmár},
  \citenamefont {Cui}, \citenamefont {Curtis}, \citenamefont {Defienne},
  \citenamefont {Hofer}, \citenamefont {Horisaki}, \citenamefont {Horstmeyer},
  \citenamefont {Ji}, \citenamefont {LaViolette}, \citenamefont {Mertz},
  \citenamefont {Moser}, \citenamefont {Mosk}, \citenamefont {Pégard},
  \citenamefont {Piestun}, \citenamefont {Popoff}, \citenamefont {Phillips},
  \citenamefont {Psaltis}, \citenamefont {Rahmani}, \citenamefont {Rigneault},
  \citenamefont {Rotter}, \citenamefont {Tian}, \citenamefont {Vellekoop},
  \citenamefont {Waller}, \citenamefont {Wang}, \citenamefont {Weber},
  \citenamefont {Xiao}, \citenamefont {Xu}, \citenamefont {Yamilov},
  \citenamefont {Yang},\ and\ \citenamefont {Yılmaz}}]{gigan_roadmap_2022}%
  \BibitemOpen
  \bibfield  {author} {\bibinfo {author} {\bibfnamefont {S.}~\bibnamefont
  {Gigan}}, \bibinfo {author} {\bibfnamefont {O.}~\bibnamefont {Katz}},
  \bibinfo {author} {\bibfnamefont {H.~B.}\ \bibnamefont {de~Aguiar}}, \bibinfo
  {author} {\bibfnamefont {E.~R.}\ \bibnamefont {Andresen}}, \bibinfo {author}
  {\bibfnamefont {A.}~\bibnamefont {Aubry}}, \bibinfo {author} {\bibfnamefont
  {J.}~\bibnamefont {Bertolotti}}, \bibinfo {author} {\bibfnamefont
  {E.}~\bibnamefont {Bossy}}, \bibinfo {author} {\bibfnamefont
  {D.}~\bibnamefont {Bouchet}}, \bibinfo {author} {\bibfnamefont
  {J.}~\bibnamefont {Brake}}, \bibinfo {author} {\bibfnamefont
  {S.}~\bibnamefont {Brasselet}}, \bibinfo {author} {\bibfnamefont
  {Y.}~\bibnamefont {Bromberg}}, \bibinfo {author} {\bibfnamefont
  {H.}~\bibnamefont {Cao}}, \bibinfo {author} {\bibfnamefont {T.}~\bibnamefont
  {Chaigne}}, \bibinfo {author} {\bibfnamefont {Z.}~\bibnamefont {Cheng}},
  \bibinfo {author} {\bibfnamefont {W.}~\bibnamefont {Choi}}, \bibinfo {author}
  {\bibfnamefont {T.}~\bibnamefont {Čižmár}}, \bibinfo {author}
  {\bibfnamefont {M.}~\bibnamefont {Cui}}, \bibinfo {author} {\bibfnamefont
  {V.~R.}\ \bibnamefont {Curtis}}, \bibinfo {author} {\bibfnamefont
  {H.}~\bibnamefont {Defienne}}, \bibinfo {author} {\bibfnamefont
  {M.}~\bibnamefont {Hofer}}, \bibinfo {author} {\bibfnamefont
  {R.}~\bibnamefont {Horisaki}}, \bibinfo {author} {\bibfnamefont
  {R.}~\bibnamefont {Horstmeyer}}, \bibinfo {author} {\bibfnamefont
  {N.}~\bibnamefont {Ji}}, \bibinfo {author} {\bibfnamefont {A.~K.}\
  \bibnamefont {LaViolette}}, \bibinfo {author} {\bibfnamefont
  {J.}~\bibnamefont {Mertz}}, \bibinfo {author} {\bibfnamefont
  {C.}~\bibnamefont {Moser}}, \bibinfo {author} {\bibfnamefont {A.~P.}\
  \bibnamefont {Mosk}}, \bibinfo {author} {\bibfnamefont {N.~C.}\ \bibnamefont
  {Pégard}}, \bibinfo {author} {\bibfnamefont {R.}~\bibnamefont {Piestun}},
  \bibinfo {author} {\bibfnamefont {S.}~\bibnamefont {Popoff}}, \bibinfo
  {author} {\bibfnamefont {D.~B.}\ \bibnamefont {Phillips}}, \bibinfo {author}
  {\bibfnamefont {D.}~\bibnamefont {Psaltis}}, \bibinfo {author} {\bibfnamefont
  {B.}~\bibnamefont {Rahmani}}, \bibinfo {author} {\bibfnamefont
  {H.}~\bibnamefont {Rigneault}}, \bibinfo {author} {\bibfnamefont
  {S.}~\bibnamefont {Rotter}}, \bibinfo {author} {\bibfnamefont
  {L.}~\bibnamefont {Tian}}, \bibinfo {author} {\bibfnamefont {I.~M.}\
  \bibnamefont {Vellekoop}}, \bibinfo {author} {\bibfnamefont {L.}~\bibnamefont
  {Waller}}, \bibinfo {author} {\bibfnamefont {L.}~\bibnamefont {Wang}},
  \bibinfo {author} {\bibfnamefont {T.}~\bibnamefont {Weber}}, \bibinfo
  {author} {\bibfnamefont {S.}~\bibnamefont {Xiao}}, \bibinfo {author}
  {\bibfnamefont {C.}~\bibnamefont {Xu}}, \bibinfo {author} {\bibfnamefont
  {A.}~\bibnamefont {Yamilov}}, \bibinfo {author} {\bibfnamefont
  {C.}~\bibnamefont {Yang}},\ and\ \bibinfo {author} {\bibfnamefont
  {H.}~\bibnamefont {Yılmaz}},\ }\href
  {https://doi.org/10.1088/2515-7647/ac76f9} {\bibfield  {journal} {\bibinfo
  {journal} {Journal of Physics: Photonics}\ }\textbf {\bibinfo {volume} {4}},\
  \bibinfo {pages} {042501} (\bibinfo {year} {2022})}\BibitemShut {NoStop}%
\bibitem [{\citenamefont {Gérardin}\ \emph {et~al.}(2014)\citenamefont
  {Gérardin}, \citenamefont {Laurent}, \citenamefont {Derode}, \citenamefont
  {Prada},\ and\ \citenamefont {Aubry}}]{gerardin_full_2014}%
  \BibitemOpen
  \bibfield  {author} {\bibinfo {author} {\bibfnamefont {B.}~\bibnamefont
  {Gérardin}}, \bibinfo {author} {\bibfnamefont {J.}~\bibnamefont {Laurent}},
  \bibinfo {author} {\bibfnamefont {A.}~\bibnamefont {Derode}}, \bibinfo
  {author} {\bibfnamefont {C.}~\bibnamefont {Prada}},\ and\ \bibinfo {author}
  {\bibfnamefont {A.}~\bibnamefont {Aubry}},\ }\href
  {https://doi.org/10.1103/PhysRevLett.113.173901} {\bibfield  {journal}
  {\bibinfo  {journal} {Physical Review Letters}\ }\textbf {\bibinfo {volume}
  {113}},\ \bibinfo {pages} {173901} (\bibinfo {year} {2014})}\BibitemShut
  {NoStop}%
\bibitem [{\citenamefont {Sarma}\ \emph {et~al.}(2016)\citenamefont {Sarma},
  \citenamefont {Yamilov}, \citenamefont {Petrenko}, \citenamefont {Bromberg},\
  and\ \citenamefont {Cao}}]{sarma_control_2016}%
  \BibitemOpen
  \bibfield  {author} {\bibinfo {author} {\bibfnamefont {R.}~\bibnamefont
  {Sarma}}, \bibinfo {author} {\bibfnamefont {A.~G.}\ \bibnamefont {Yamilov}},
  \bibinfo {author} {\bibfnamefont {S.}~\bibnamefont {Petrenko}}, \bibinfo
  {author} {\bibfnamefont {Y.}~\bibnamefont {Bromberg}},\ and\ \bibinfo
  {author} {\bibfnamefont {H.}~\bibnamefont {Cao}},\ }\href
  {https://doi.org/10.1103/PhysRevLett.117.086803} {\bibfield  {journal}
  {\bibinfo  {journal} {Physical Review Letters}\ }\textbf {\bibinfo {volume}
  {117}},\ \bibinfo {pages} {086803} (\bibinfo {year} {2016})}\BibitemShut
  {NoStop}%
\bibitem [{\citenamefont {Horstmeyer}\ \emph {et~al.}(2015)\citenamefont
  {Horstmeyer}, \citenamefont {Ruan},\ and\ \citenamefont
  {Yang}}]{horstmeyer_guidestar-assisted_2015}%
  \BibitemOpen
  \bibfield  {author} {\bibinfo {author} {\bibfnamefont {R.}~\bibnamefont
  {Horstmeyer}}, \bibinfo {author} {\bibfnamefont {H.}~\bibnamefont {Ruan}},\
  and\ \bibinfo {author} {\bibfnamefont {C.}~\bibnamefont {Yang}},\ }\href
  {https://doi.org/10.1038/nphoton.2015.140} {\bibfield  {journal} {\bibinfo
  {journal} {Nature Photonics}\ }\textbf {\bibinfo {volume} {9}},\ \bibinfo
  {pages} {563} (\bibinfo {year} {2015})}\BibitemShut {NoStop}%
\bibitem [{\citenamefont {Vellekoop}\ and\ \citenamefont
  {Mosk}(2007)}]{vellekoop_focusing_2007}%
  \BibitemOpen
  \bibfield  {author} {\bibinfo {author} {\bibfnamefont {I.~M.}\ \bibnamefont
  {Vellekoop}}\ and\ \bibinfo {author} {\bibfnamefont {A.~P.}\ \bibnamefont
  {Mosk}},\ }\href {https://doi.org/10.1364/OL.32.002309} {\bibfield  {journal}
  {\bibinfo  {journal} {Optics Letters}\ }\textbf {\bibinfo {volume} {32}},\
  \bibinfo {pages} {2309} (\bibinfo {year} {2007})}\BibitemShut {NoStop}%
\bibitem [{\citenamefont {Efremidis}\ and\ \citenamefont
  {Christodoulides}(2010)}]{efremidis_abruptly_2010}%
  \BibitemOpen
  \bibfield  {author} {\bibinfo {author} {\bibfnamefont {N.~K.}\ \bibnamefont
  {Efremidis}}\ and\ \bibinfo {author} {\bibfnamefont {D.~N.}\ \bibnamefont
  {Christodoulides}},\ }\href {https://doi.org/10.1364/OL.35.004045} {\bibfield
   {journal} {\bibinfo  {journal} {Optics Letters}\ }\textbf {\bibinfo {volume}
  {35}},\ \bibinfo {pages} {4045} (\bibinfo {year} {2010})}\BibitemShut
  {NoStop}%
\bibitem [{\citenamefont {Ashkin}\ \emph {et~al.}(1986)\citenamefont {Ashkin},
  \citenamefont {Dziedzic}, \citenamefont {Bjorkholm},\ and\ \citenamefont
  {Chu}}]{ashkin_observation_1986}%
  \BibitemOpen
  \bibfield  {author} {\bibinfo {author} {\bibfnamefont {A.}~\bibnamefont
  {Ashkin}}, \bibinfo {author} {\bibfnamefont {J.~M.}\ \bibnamefont
  {Dziedzic}}, \bibinfo {author} {\bibfnamefont {J.~E.}\ \bibnamefont
  {Bjorkholm}},\ and\ \bibinfo {author} {\bibfnamefont {S.}~\bibnamefont
  {Chu}},\ }\href {https://doi.org/10.1364/OL.11.000288} {\bibfield  {journal}
  {\bibinfo  {journal} {Optics Letters}\ }\textbf {\bibinfo {volume} {11}},\
  \bibinfo {pages} {288} (\bibinfo {year} {1986})}\BibitemShut {NoStop}%
\bibitem [{\citenamefont {Block}\ \emph {et~al.}(1990)\citenamefont {Block},
  \citenamefont {Goldstein},\ and\ \citenamefont {Schnapp}}]{block_bead_1990}%
  \BibitemOpen
  \bibfield  {author} {\bibinfo {author} {\bibfnamefont {S.~M.}\ \bibnamefont
  {Block}}, \bibinfo {author} {\bibfnamefont {L.~S.~B.}\ \bibnamefont
  {Goldstein}},\ and\ \bibinfo {author} {\bibfnamefont {B.~J.}\ \bibnamefont
  {Schnapp}},\ }\href {https://doi.org/10.1038/348348a0} {\bibfield  {journal}
  {\bibinfo  {journal} {Nature}\ }\textbf {\bibinfo {volume} {348}},\ \bibinfo
  {pages} {348} (\bibinfo {year} {1990})}\BibitemShut {NoStop}%
\bibitem [{\citenamefont {Būtaitė}\ \emph {et~al.}(2019)\citenamefont
  {Būtaitė}, \citenamefont {Gibson}, \citenamefont {Ho}, \citenamefont
  {Taverne}, \citenamefont {Taylor},\ and\ \citenamefont
  {Phillips}}]{butaite_indirect_2019}%
  \BibitemOpen
  \bibfield  {author} {\bibinfo {author} {\bibfnamefont {U.~G.}\ \bibnamefont
  {Būtaitė}}, \bibinfo {author} {\bibfnamefont {G.~M.}\ \bibnamefont
  {Gibson}}, \bibinfo {author} {\bibfnamefont {Y.-L.~D.}\ \bibnamefont {Ho}},
  \bibinfo {author} {\bibfnamefont {M.}~\bibnamefont {Taverne}}, \bibinfo
  {author} {\bibfnamefont {J.~M.}\ \bibnamefont {Taylor}},\ and\ \bibinfo
  {author} {\bibfnamefont {D.~B.}\ \bibnamefont {Phillips}},\ }\href
  {https://doi.org/10.1038/s41467-019-08968-7} {\bibfield  {journal} {\bibinfo
  {journal} {Nature Communications}\ }\textbf {\bibinfo {volume} {10}},\
  \bibinfo {pages} {1215} (\bibinfo {year} {2019})}\BibitemShut {NoStop}%
\bibitem [{\citenamefont {Magrini}\ \emph {et~al.}(2021)\citenamefont
  {Magrini}, \citenamefont {Rosenzweig}, \citenamefont {Bach}, \citenamefont
  {Deutschmann-Olek}, \citenamefont {Hofer}, \citenamefont {Hong},
  \citenamefont {Kiesel}, \citenamefont {Kugi},\ and\ \citenamefont
  {Aspelmeyer}}]{magrini_real-time_2021}%
  \BibitemOpen
  \bibfield  {author} {\bibinfo {author} {\bibfnamefont {L.}~\bibnamefont
  {Magrini}}, \bibinfo {author} {\bibfnamefont {P.}~\bibnamefont {Rosenzweig}},
  \bibinfo {author} {\bibfnamefont {C.}~\bibnamefont {Bach}}, \bibinfo {author}
  {\bibfnamefont {A.}~\bibnamefont {Deutschmann-Olek}}, \bibinfo {author}
  {\bibfnamefont {S.~G.}\ \bibnamefont {Hofer}}, \bibinfo {author}
  {\bibfnamefont {S.}~\bibnamefont {Hong}}, \bibinfo {author} {\bibfnamefont
  {N.}~\bibnamefont {Kiesel}}, \bibinfo {author} {\bibfnamefont
  {A.}~\bibnamefont {Kugi}},\ and\ \bibinfo {author} {\bibfnamefont
  {M.}~\bibnamefont {Aspelmeyer}},\ }\href
  {https://doi.org/10.1038/s41586-021-03602-3} {\bibfield  {journal} {\bibinfo
  {journal} {Nature}\ }\textbf {\bibinfo {volume} {595}},\ \bibinfo {pages}
  {373} (\bibinfo {year} {2021})}\BibitemShut {NoStop}%
\bibitem [{\citenamefont {Hüpfl}\ \emph {et~al.}(2023)\citenamefont {Hüpfl},
  \citenamefont {Bachelard}, \citenamefont {Kaczvinszki}, \citenamefont
  {Horodynski}, \citenamefont {Kühmayer},\ and\ \citenamefont
  {Rotter}}]{hupfl_optimal_2023}%
  \BibitemOpen
  \bibfield  {author} {\bibinfo {author} {\bibfnamefont {J.}~\bibnamefont
  {Hüpfl}}, \bibinfo {author} {\bibfnamefont {N.}~\bibnamefont {Bachelard}},
  \bibinfo {author} {\bibfnamefont {M.}~\bibnamefont {Kaczvinszki}}, \bibinfo
  {author} {\bibfnamefont {M.}~\bibnamefont {Horodynski}}, \bibinfo {author}
  {\bibfnamefont {M.}~\bibnamefont {Kühmayer}},\ and\ \bibinfo {author}
  {\bibfnamefont {S.}~\bibnamefont {Rotter}},\ }\href
  {https://doi.org/10.1103/PhysRevA.107.023112} {\bibfield  {journal} {\bibinfo
   {journal} {Physical Review A}\ }\textbf {\bibinfo {volume} {107}},\ \bibinfo
  {pages} {023112} (\bibinfo {year} {2023})}\BibitemShut {NoStop}%
\bibitem [{\citenamefont {Hirayama}\ \emph {et~al.}(2019)\citenamefont
  {Hirayama}, \citenamefont {Martinez~Plasencia}, \citenamefont {Masuda},\ and\
  \citenamefont {Subramanian}}]{hirayama_volumetric_2019}%
  \BibitemOpen
  \bibfield  {author} {\bibinfo {author} {\bibfnamefont {R.}~\bibnamefont
  {Hirayama}}, \bibinfo {author} {\bibfnamefont {D.}~\bibnamefont
  {Martinez~Plasencia}}, \bibinfo {author} {\bibfnamefont {N.}~\bibnamefont
  {Masuda}},\ and\ \bibinfo {author} {\bibfnamefont {S.}~\bibnamefont
  {Subramanian}},\ }\href {https://doi.org/10.1038/s41586-019-1739-5}
  {\bibfield  {journal} {\bibinfo  {journal} {Nature}\ }\textbf {\bibinfo
  {volume} {575}},\ \bibinfo {pages} {320} (\bibinfo {year}
  {2019})}\BibitemShut {NoStop}%
\bibitem [{\citenamefont {Ghanem}\ \emph {et~al.}(2020)\citenamefont {Ghanem},
  \citenamefont {Maxwell}, \citenamefont {Wang}, \citenamefont {Cunitz},
  \citenamefont {Khokhlova}, \citenamefont {Sapozhnikov},\ and\ \citenamefont
  {Bailey}}]{ghanem_noninvasive_2020}%
  \BibitemOpen
  \bibfield  {author} {\bibinfo {author} {\bibfnamefont {M.~A.}\ \bibnamefont
  {Ghanem}}, \bibinfo {author} {\bibfnamefont {A.~D.}\ \bibnamefont {Maxwell}},
  \bibinfo {author} {\bibfnamefont {Y.-N.}\ \bibnamefont {Wang}}, \bibinfo
  {author} {\bibfnamefont {B.~W.}\ \bibnamefont {Cunitz}}, \bibinfo {author}
  {\bibfnamefont {V.~A.}\ \bibnamefont {Khokhlova}}, \bibinfo {author}
  {\bibfnamefont {O.~A.}\ \bibnamefont {Sapozhnikov}},\ and\ \bibinfo {author}
  {\bibfnamefont {M.~R.}\ \bibnamefont {Bailey}},\ }\href
  {https://doi.org/10.1073/pnas.2001779117} {\bibfield  {journal} {\bibinfo
  {journal} {Proceedings of the National Academy of Sciences}\ }\textbf
  {\bibinfo {volume} {117}},\ \bibinfo {pages} {16848} (\bibinfo {year}
  {2020})}\BibitemShut {NoStop}%
\bibitem [{\citenamefont {Chen}\ \emph {et~al.}(2011)\citenamefont {Chen},
  \citenamefont {Ng}, \citenamefont {Lin},\ and\ \citenamefont
  {Chan}}]{chen_optical_2011}%
  \BibitemOpen
  \bibfield  {author} {\bibinfo {author} {\bibfnamefont {J.}~\bibnamefont
  {Chen}}, \bibinfo {author} {\bibfnamefont {J.}~\bibnamefont {Ng}}, \bibinfo
  {author} {\bibfnamefont {Z.}~\bibnamefont {Lin}},\ and\ \bibinfo {author}
  {\bibfnamefont {C.~T.}\ \bibnamefont {Chan}},\ }\href
  {https://doi.org/10.1038/nphoton.2011.153} {\bibfield  {journal} {\bibinfo
  {journal} {Nature Photonics}\ }\textbf {\bibinfo {volume} {5}},\ \bibinfo
  {pages} {531} (\bibinfo {year} {2011})}\BibitemShut {NoStop}%
\bibitem [{\citenamefont {Novitsky}\ \emph {et~al.}(2011)\citenamefont
  {Novitsky}, \citenamefont {Qiu},\ and\ \citenamefont
  {Wang}}]{novitsky_single_2011}%
  \BibitemOpen
  \bibfield  {author} {\bibinfo {author} {\bibfnamefont {A.}~\bibnamefont
  {Novitsky}}, \bibinfo {author} {\bibfnamefont {C.-W.}\ \bibnamefont {Qiu}},\
  and\ \bibinfo {author} {\bibfnamefont {H.}~\bibnamefont {Wang}},\ }\href
  {https://doi.org/10.1103/PhysRevLett.107.203601} {\bibfield  {journal}
  {\bibinfo  {journal} {Physical Review Letters}\ }\textbf {\bibinfo {volume}
  {107}},\ \bibinfo {pages} {203601} (\bibinfo {year} {2011})}\BibitemShut
  {NoStop}%
\bibitem [{\citenamefont {Brzobohatý}\ \emph {et~al.}(2013)\citenamefont
  {Brzobohatý}, \citenamefont {Karásek}, \citenamefont {Šiler},
  \citenamefont {Chvátal}, \citenamefont {Čižmár},\ and\ \citenamefont
  {Zemánek}}]{brzobohaty_experimental_2013}%
  \BibitemOpen
  \bibfield  {author} {\bibinfo {author} {\bibfnamefont {O.}~\bibnamefont
  {Brzobohatý}}, \bibinfo {author} {\bibfnamefont {V.}~\bibnamefont
  {Karásek}}, \bibinfo {author} {\bibfnamefont {M.}~\bibnamefont {Šiler}},
  \bibinfo {author} {\bibfnamefont {L.}~\bibnamefont {Chvátal}}, \bibinfo
  {author} {\bibfnamefont {T.}~\bibnamefont {Čižmár}},\ and\ \bibinfo
  {author} {\bibfnamefont {P.}~\bibnamefont {Zemánek}},\ }\href
  {https://doi.org/10.1038/nphoton.2012.332} {\bibfield  {journal} {\bibinfo
  {journal} {Nature Photonics}\ }\textbf {\bibinfo {volume} {7}},\ \bibinfo
  {pages} {123} (\bibinfo {year} {2013})}\BibitemShut {NoStop}%
\bibitem [{\citenamefont {Li}\ \emph {et~al.}(2020)\citenamefont {Li},
  \citenamefont {Cao}, \citenamefont {Zhou}, \citenamefont {Xu}, \citenamefont
  {Zhu}, \citenamefont {Shi}, \citenamefont {Qiu},\ and\ \citenamefont
  {Ding}}]{li_optical_2020}%
  \BibitemOpen
  \bibfield  {author} {\bibinfo {author} {\bibfnamefont {H.}~\bibnamefont
  {Li}}, \bibinfo {author} {\bibfnamefont {Y.}~\bibnamefont {Cao}}, \bibinfo
  {author} {\bibfnamefont {L.-M.}\ \bibnamefont {Zhou}}, \bibinfo {author}
  {\bibfnamefont {X.}~\bibnamefont {Xu}}, \bibinfo {author} {\bibfnamefont
  {T.}~\bibnamefont {Zhu}}, \bibinfo {author} {\bibfnamefont {Y.}~\bibnamefont
  {Shi}}, \bibinfo {author} {\bibfnamefont {C.-W.}\ \bibnamefont {Qiu}},\ and\
  \bibinfo {author} {\bibfnamefont {W.}~\bibnamefont {Ding}},\ }\href
  {https://doi.org/10.1364/AOP.378390} {\bibfield  {journal} {\bibinfo
  {journal} {Advances in Optics and Photonics}\ }\textbf {\bibinfo {volume}
  {12}},\ \bibinfo {pages} {288} (\bibinfo {year} {2020})}\BibitemShut
  {NoStop}%
\bibitem [{\citenamefont {Yevick}\ \emph {et~al.}(2016)\citenamefont {Yevick},
  \citenamefont {Ruffner},\ and\ \citenamefont {Grier}}]{yevick_tractor_2016}%
  \BibitemOpen
  \bibfield  {author} {\bibinfo {author} {\bibfnamefont {A.}~\bibnamefont
  {Yevick}}, \bibinfo {author} {\bibfnamefont {D.~B.}\ \bibnamefont
  {Ruffner}},\ and\ \bibinfo {author} {\bibfnamefont {D.~G.}\ \bibnamefont
  {Grier}},\ }\href {https://doi.org/10.1103/PhysRevA.93.043807} {\bibfield
  {journal} {\bibinfo  {journal} {Physical Review A}\ }\textbf {\bibinfo
  {volume} {93}},\ \bibinfo {pages} {043807} (\bibinfo {year}
  {2016})}\BibitemShut {NoStop}%
\bibitem [{\citenamefont {Abdelaziz}\ and\ \citenamefont
  {Grier}(2020)}]{abdelaziz_acoustokinetics_2020}%
  \BibitemOpen
  \bibfield  {author} {\bibinfo {author} {\bibfnamefont {M.~A.}\ \bibnamefont
  {Abdelaziz}}\ and\ \bibinfo {author} {\bibfnamefont {D.~G.}\ \bibnamefont
  {Grier}},\ }\href {https://doi.org/10.1103/PhysRevResearch.2.013172}
  {\bibfield  {journal} {\bibinfo  {journal} {Physical Review Research}\
  }\textbf {\bibinfo {volume} {2}},\ \bibinfo {pages} {013172} (\bibinfo {year}
  {2020})}\BibitemShut {NoStop}%
\bibitem [{\citenamefont {Marzo}\ \emph {et~al.}(2015)\citenamefont {Marzo},
  \citenamefont {Seah}, \citenamefont {Drinkwater}, \citenamefont {Sahoo},
  \citenamefont {Long},\ and\ \citenamefont
  {Subramanian}}]{marzo_holographic_2015}%
  \BibitemOpen
  \bibfield  {author} {\bibinfo {author} {\bibfnamefont {A.}~\bibnamefont
  {Marzo}}, \bibinfo {author} {\bibfnamefont {S.~A.}\ \bibnamefont {Seah}},
  \bibinfo {author} {\bibfnamefont {B.~W.}\ \bibnamefont {Drinkwater}},
  \bibinfo {author} {\bibfnamefont {D.~R.}\ \bibnamefont {Sahoo}}, \bibinfo
  {author} {\bibfnamefont {B.}~\bibnamefont {Long}},\ and\ \bibinfo {author}
  {\bibfnamefont {S.}~\bibnamefont {Subramanian}},\ }\href
  {https://doi.org/10.1038/ncomms9661} {\bibfield  {journal} {\bibinfo
  {journal} {Nature Communications}\ }\textbf {\bibinfo {volume} {6}},\
  \bibinfo {pages} {8661} (\bibinfo {year} {2015})}\BibitemShut {NoStop}%
\bibitem [{\citenamefont {Démoré}\ \emph {et~al.}(2014)\citenamefont
  {Démoré}, \citenamefont {Dahl}, \citenamefont {Yang}, \citenamefont
  {Glynne-Jones}, \citenamefont {Melzer}, \citenamefont {Cochran},
  \citenamefont {MacDonald},\ and\ \citenamefont
  {Spalding}}]{demore_acoustic_2014}%
  \BibitemOpen
  \bibfield  {author} {\bibinfo {author} {\bibfnamefont {C.~E.}\ \bibnamefont
  {Démoré}}, \bibinfo {author} {\bibfnamefont {P.~M.}\ \bibnamefont {Dahl}},
  \bibinfo {author} {\bibfnamefont {Z.}~\bibnamefont {Yang}}, \bibinfo {author}
  {\bibfnamefont {P.}~\bibnamefont {Glynne-Jones}}, \bibinfo {author}
  {\bibfnamefont {A.}~\bibnamefont {Melzer}}, \bibinfo {author} {\bibfnamefont
  {S.}~\bibnamefont {Cochran}}, \bibinfo {author} {\bibfnamefont
  {M.}~\bibnamefont {MacDonald}},\ and\ \bibinfo {author} {\bibfnamefont
  {G.~C.}\ \bibnamefont {Spalding}},\ }\href
  {https://doi.org/10.1103/PhysRevLett.112.174302} {\bibfield  {journal}
  {\bibinfo  {journal} {Physical Review Letters}\ }\textbf {\bibinfo {volume}
  {112}},\ \bibinfo {pages} {174302} (\bibinfo {year} {2014})}\BibitemShut
  {NoStop}%
\bibitem [{\citenamefont {Shvedov}\ \emph {et~al.}(2014)\citenamefont
  {Shvedov}, \citenamefont {Davoyan}, \citenamefont {Hnatovsky}, \citenamefont
  {Engheta},\ and\ \citenamefont {Krolikowski}}]{shvedov_long-range_2014}%
  \BibitemOpen
  \bibfield  {author} {\bibinfo {author} {\bibfnamefont {V.}~\bibnamefont
  {Shvedov}}, \bibinfo {author} {\bibfnamefont {A.~R.}\ \bibnamefont
  {Davoyan}}, \bibinfo {author} {\bibfnamefont {C.}~\bibnamefont {Hnatovsky}},
  \bibinfo {author} {\bibfnamefont {N.}~\bibnamefont {Engheta}},\ and\ \bibinfo
  {author} {\bibfnamefont {W.}~\bibnamefont {Krolikowski}},\ }\href
  {https://doi.org/10.1038/nphoton.2014.242} {\bibfield  {journal} {\bibinfo
  {journal} {Nature Photonics}\ }\textbf {\bibinfo {volume} {8}},\ \bibinfo
  {pages} {846} (\bibinfo {year} {2014})}\BibitemShut {NoStop}%
\bibitem [{\citenamefont {Li}\ \emph {et~al.}(2019)\citenamefont {Li},
  \citenamefont {Chen}, \citenamefont {Lin},\ and\ \citenamefont
  {Ng}}]{li_optical_2019}%
  \BibitemOpen
  \bibfield  {author} {\bibinfo {author} {\bibfnamefont {X.}~\bibnamefont
  {Li}}, \bibinfo {author} {\bibfnamefont {J.}~\bibnamefont {Chen}}, \bibinfo
  {author} {\bibfnamefont {Z.}~\bibnamefont {Lin}},\ and\ \bibinfo {author}
  {\bibfnamefont {J.}~\bibnamefont {Ng}},\ }\href
  {https://doi.org/10.1126/sciadv.aau7814} {\bibfield  {journal} {\bibinfo
  {journal} {Science Advances}\ }\textbf {\bibinfo {volume} {5}},\ \bibinfo
  {pages} {eaau7814} (\bibinfo {year} {2019})}\BibitemShut {NoStop}%
\bibitem [{\citenamefont {Ding}\ \emph {et~al.}(2014)\citenamefont {Ding},
  \citenamefont {Ng}, \citenamefont {Zhou},\ and\ \citenamefont
  {Chan}}]{ding_realization_2014}%
  \BibitemOpen
  \bibfield  {author} {\bibinfo {author} {\bibfnamefont {K.}~\bibnamefont
  {Ding}}, \bibinfo {author} {\bibfnamefont {J.}~\bibnamefont {Ng}}, \bibinfo
  {author} {\bibfnamefont {L.}~\bibnamefont {Zhou}},\ and\ \bibinfo {author}
  {\bibfnamefont {C.~T.}\ \bibnamefont {Chan}},\ }\href
  {https://doi.org/10.1103/PhysRevA.89.063825} {\bibfield  {journal} {\bibinfo
  {journal} {Physical Review A}\ }\textbf {\bibinfo {volume} {89}},\ \bibinfo
  {pages} {063825} (\bibinfo {year} {2014})}\BibitemShut {NoStop}%
\bibitem [{\citenamefont {Fernandes}\ and\ \citenamefont
  {Silveirinha}(2015)}]{fernandes_optical_2015}%
  \BibitemOpen
  \bibfield  {author} {\bibinfo {author} {\bibfnamefont {D.~E.}\ \bibnamefont
  {Fernandes}}\ and\ \bibinfo {author} {\bibfnamefont {M.~G.}\ \bibnamefont
  {Silveirinha}},\ }\href {https://doi.org/10.1103/PhysRevA.91.061801}
  {\bibfield  {journal} {\bibinfo  {journal} {Physical Review A}\ }\textbf
  {\bibinfo {volume} {91}},\ \bibinfo {pages} {061801} (\bibinfo {year}
  {2015})}\BibitemShut {NoStop}%
\bibitem [{\citenamefont {Kuang}\ \emph {et~al.}(2020)\citenamefont {Kuang},
  \citenamefont {Zhang},\ and\ \citenamefont {Miller}}]{kuang_maximal_2020}%
  \BibitemOpen
  \bibfield  {author} {\bibinfo {author} {\bibfnamefont {Z.}~\bibnamefont
  {Kuang}}, \bibinfo {author} {\bibfnamefont {L.}~\bibnamefont {Zhang}},\ and\
  \bibinfo {author} {\bibfnamefont {O.~D.}\ \bibnamefont {Miller}},\ }\href
  {https://doi.org/10.1364/OPTICA.398715} {\bibfield  {journal} {\bibinfo
  {journal} {Optica}\ }\textbf {\bibinfo {volume} {7}},\ \bibinfo {pages}
  {1746} (\bibinfo {year} {2020})}\BibitemShut {NoStop}%
\bibitem [{\citenamefont {Liu}\ \emph {et~al.}(2019)\citenamefont {Liu},
  \citenamefont {Fan}, \citenamefont {Lee}, \citenamefont {Fang}, \citenamefont
  {Johnson},\ and\ \citenamefont {Miller}}]{liu_optimal_2019}%
  \BibitemOpen
  \bibfield  {author} {\bibinfo {author} {\bibfnamefont {Y.}~\bibnamefont
  {Liu}}, \bibinfo {author} {\bibfnamefont {L.}~\bibnamefont {Fan}}, \bibinfo
  {author} {\bibfnamefont {Y.~E.}\ \bibnamefont {Lee}}, \bibinfo {author}
  {\bibfnamefont {N.~X.}\ \bibnamefont {Fang}}, \bibinfo {author}
  {\bibfnamefont {S.~G.}\ \bibnamefont {Johnson}},\ and\ \bibinfo {author}
  {\bibfnamefont {O.~D.}\ \bibnamefont {Miller}},\ }\href
  {https://doi.org/10.1021/acsphotonics.8b01263} {\bibfield  {journal}
  {\bibinfo  {journal} {ACS Photonics}\ }\textbf {\bibinfo {volume} {6}},\
  \bibinfo {pages} {395} (\bibinfo {year} {2019})}\BibitemShut {NoStop}%
\bibitem [{\citenamefont {Ambichl}\ \emph {et~al.}(2017)\citenamefont
  {Ambichl}, \citenamefont {Brandstötter}, \citenamefont {Böhm},
  \citenamefont {Kühmayer}, \citenamefont {Kuhl},\ and\ \citenamefont
  {Rotter}}]{ambichl_focusing_2017}%
  \BibitemOpen
  \bibfield  {author} {\bibinfo {author} {\bibfnamefont {P.}~\bibnamefont
  {Ambichl}}, \bibinfo {author} {\bibfnamefont {A.}~\bibnamefont
  {Brandstötter}}, \bibinfo {author} {\bibfnamefont {J.}~\bibnamefont
  {Böhm}}, \bibinfo {author} {\bibfnamefont {M.}~\bibnamefont {Kühmayer}},
  \bibinfo {author} {\bibfnamefont {U.}~\bibnamefont {Kuhl}},\ and\ \bibinfo
  {author} {\bibfnamefont {S.}~\bibnamefont {Rotter}},\ }\href
  {https://doi.org/10.1103/PhysRevLett.119.033903} {\bibfield  {journal}
  {\bibinfo  {journal} {Physical Review Letters}\ }\textbf {\bibinfo {volume}
  {119}},\ \bibinfo {pages} {033903} (\bibinfo {year} {2017})}\BibitemShut
  {NoStop}%
\bibitem [{\citenamefont {Horodynski}\ \emph {et~al.}(2020)\citenamefont
  {Horodynski}, \citenamefont {Kühmayer}, \citenamefont {Brandstötter},
  \citenamefont {Pichler}, \citenamefont {Fyodorov}, \citenamefont {Kuhl},\
  and\ \citenamefont {Rotter}}]{horodynski_optimal_2020}%
  \BibitemOpen
  \bibfield  {author} {\bibinfo {author} {\bibfnamefont {M.}~\bibnamefont
  {Horodynski}}, \bibinfo {author} {\bibfnamefont {M.}~\bibnamefont
  {Kühmayer}}, \bibinfo {author} {\bibfnamefont {A.}~\bibnamefont
  {Brandstötter}}, \bibinfo {author} {\bibfnamefont {K.}~\bibnamefont
  {Pichler}}, \bibinfo {author} {\bibfnamefont {Y.~V.}\ \bibnamefont
  {Fyodorov}}, \bibinfo {author} {\bibfnamefont {U.}~\bibnamefont {Kuhl}},\
  and\ \bibinfo {author} {\bibfnamefont {S.}~\bibnamefont {Rotter}},\ }\href
  {https://doi.org/10.1038/s41566-019-0550-z} {\bibfield  {journal} {\bibinfo
  {journal} {Nature Photonics}\ }\textbf {\bibinfo {volume} {14}},\ \bibinfo
  {pages} {149} (\bibinfo {year} {2020})}\BibitemShut {NoStop}%
\bibitem [{\citenamefont {Būtaitė}\ \emph {et~al.}(2023)\citenamefont
  {Būtaitė}, \citenamefont {Sharp}, \citenamefont {Horodynski}, \citenamefont
  {Gibson}, \citenamefont {Padgett}, \citenamefont {Rotter}, \citenamefont
  {Taylor},\ and\ \citenamefont {Phillips}}]{butaite_photon-efficient_2023}%
  \BibitemOpen
  \bibfield  {author} {\bibinfo {author} {\bibfnamefont {U.~G.}\ \bibnamefont
  {Būtaitė}}, \bibinfo {author} {\bibfnamefont {C.}~\bibnamefont {Sharp}},
  \bibinfo {author} {\bibfnamefont {M.}~\bibnamefont {Horodynski}}, \bibinfo
  {author} {\bibfnamefont {G.~M.}\ \bibnamefont {Gibson}}, \bibinfo {author}
  {\bibfnamefont {M.~J.}\ \bibnamefont {Padgett}}, \bibinfo {author}
  {\bibfnamefont {S.}~\bibnamefont {Rotter}}, \bibinfo {author} {\bibfnamefont
  {J.~M.}\ \bibnamefont {Taylor}},\ and\ \bibinfo {author} {\bibfnamefont
  {D.~B.}\ \bibnamefont {Phillips}},\ }\href {http://arxiv.org/abs/2304.12848}
  {\bibinfo {title} {Photon-efficient optical tweezers via wavefront shaping}}
  (\bibinfo {year} {2023}),\ \bibinfo {note} {arXiv:2304.12848
  [physics]}\BibitemShut {NoStop}%
\bibitem [{\citenamefont {Bouchet}\ \emph {et~al.}(2021)\citenamefont
  {Bouchet}, \citenamefont {Rotter},\ and\ \citenamefont
  {Mosk}}]{bouchet_maximum_2021}%
  \BibitemOpen
  \bibfield  {author} {\bibinfo {author} {\bibfnamefont {D.}~\bibnamefont
  {Bouchet}}, \bibinfo {author} {\bibfnamefont {S.}~\bibnamefont {Rotter}},\
  and\ \bibinfo {author} {\bibfnamefont {A.~P.}\ \bibnamefont {Mosk}},\ }\href
  {https://doi.org/10.1038/s41567-020-01137-4} {\bibfield  {journal} {\bibinfo
  {journal} {Nature Physics}\ }\textbf {\bibinfo {volume} {17}},\ \bibinfo
  {pages} {564} (\bibinfo {year} {2021})}\BibitemShut {NoStop}%
\bibitem [{\citenamefont {Horodynski}\ \emph {et~al.}(2022)\citenamefont
  {Horodynski}, \citenamefont {Kühmayer}, \citenamefont {Ferise},
  \citenamefont {Rotter},\ and\ \citenamefont
  {Davy}}]{horodynski_anti-reflection_2022}%
  \BibitemOpen
  \bibfield  {author} {\bibinfo {author} {\bibfnamefont {M.}~\bibnamefont
  {Horodynski}}, \bibinfo {author} {\bibfnamefont {M.}~\bibnamefont
  {Kühmayer}}, \bibinfo {author} {\bibfnamefont {C.}~\bibnamefont {Ferise}},
  \bibinfo {author} {\bibfnamefont {S.}~\bibnamefont {Rotter}},\ and\ \bibinfo
  {author} {\bibfnamefont {M.}~\bibnamefont {Davy}},\ }\href
  {https://doi.org/10.1038/s41586-022-04843-6} {\bibfield  {journal} {\bibinfo
  {journal} {Nature}\ }\textbf {\bibinfo {volume} {607}},\ \bibinfo {pages}
  {281} (\bibinfo {year} {2022})}\BibitemShut {NoStop}%
\bibitem [{\citenamefont {Strasser}\ \emph {et~al.}(2021)\citenamefont
  {Strasser}, \citenamefont {Moser}, \citenamefont {Ritsch-Marte},\ and\
  \citenamefont {Thalhammer}}]{strasser_direct_2021}%
  \BibitemOpen
  \bibfield  {author} {\bibinfo {author} {\bibfnamefont {F.}~\bibnamefont
  {Strasser}}, \bibinfo {author} {\bibfnamefont {S.}~\bibnamefont {Moser}},
  \bibinfo {author} {\bibfnamefont {M.}~\bibnamefont {Ritsch-Marte}},\ and\
  \bibinfo {author} {\bibfnamefont {G.}~\bibnamefont {Thalhammer}},\ }\href
  {https://doi.org/10.1364/OPTICA.410494} {\bibfield  {journal} {\bibinfo
  {journal} {Optica}\ }\textbf {\bibinfo {volume} {8}},\ \bibinfo {pages} {79}
  (\bibinfo {year} {2021})}\BibitemShut {NoStop}%
\bibitem [{\citenamefont {Mazilu}\ \emph {et~al.}(2011)\citenamefont {Mazilu},
  \citenamefont {Baumgartl}, \citenamefont {Kosmeier},\ and\ \citenamefont
  {Dholakia}}]{mazilu_optical_2011}%
  \BibitemOpen
  \bibfield  {author} {\bibinfo {author} {\bibfnamefont {M.}~\bibnamefont
  {Mazilu}}, \bibinfo {author} {\bibfnamefont {J.}~\bibnamefont {Baumgartl}},
  \bibinfo {author} {\bibfnamefont {S.}~\bibnamefont {Kosmeier}},\ and\
  \bibinfo {author} {\bibfnamefont {K.}~\bibnamefont {Dholakia}},\ }\href
  {https://doi.org/10.1364/OE.19.000933} {\bibfield  {journal} {\bibinfo
  {journal} {Optics Express}\ }\textbf {\bibinfo {volume} {19}},\ \bibinfo
  {pages} {933} (\bibinfo {year} {2011})}\BibitemShut {NoStop}%
\bibitem [{\citenamefont {Horodynski}\ \emph {et~al.}(2023)\citenamefont
  {Horodynski}, \citenamefont {Reiter}, \citenamefont {Kühmayer},\ and\
  \citenamefont {Rotter}}]{horodynski_open_2023}%
  \BibitemOpen
  \bibfield  {author} {\bibinfo {author} {\bibfnamefont {M.}~\bibnamefont
  {Horodynski}}, \bibinfo {author} {\bibfnamefont {T.}~\bibnamefont {Reiter}},
  \bibinfo {author} {\bibfnamefont {M.}~\bibnamefont {Kühmayer}},\ and\
  \bibinfo {author} {\bibfnamefont {S.}~\bibnamefont {Rotter}},\ }\href
  {https://github.com/michaelhorodynski/Open-Scattering-Systems} {\bibinfo
  {title} {Open {Scattering} {Systems}}} (\bibinfo {year} {2023})\BibitemShut
  {NoStop}%
\bibitem [{\citenamefont {Schöberl}(1997)}]{schoberl_netgen_1997}%
  \BibitemOpen
  \bibfield  {author} {\bibinfo {author} {\bibfnamefont {J.}~\bibnamefont
  {Schöberl}},\ }\href {https://doi.org/10.1007/s007910050004} {\bibfield
  {journal} {\bibinfo  {journal} {Computing and Visualization in Science}\
  }\textbf {\bibinfo {volume} {1}},\ \bibinfo {pages} {41} (\bibinfo {year}
  {1997})}\BibitemShut {NoStop}%
\bibitem [{\citenamefont {Schöberl}(2014)}]{schoberl_c11_2014}%
  \BibitemOpen
  \bibfield  {author} {\bibinfo {author} {\bibfnamefont {J.}~\bibnamefont
  {Schöberl}},\ }\href@noop {} {\emph {\bibinfo {title} {C++11
  {Implementation} of {Finite} {Elements} in {NGSolve}}}},\ \bibinfo {type}
  {{ASC} {Report}}\ (\bibinfo  {institution} {Institute for Analysis and
  Scientific Computing, Vienna University of Technology},\ \bibinfo {year}
  {2014})\BibitemShut {NoStop}%
\bibitem [{\citenamefont {Sukhov}\ and\ \citenamefont
  {Dogariu}(2011)}]{sukhov_negative_2011}%
  \BibitemOpen
  \bibfield  {author} {\bibinfo {author} {\bibfnamefont {S.}~\bibnamefont
  {Sukhov}}\ and\ \bibinfo {author} {\bibfnamefont {A.}~\bibnamefont
  {Dogariu}},\ }\href {https://doi.org/10.1103/PhysRevLett.107.203602}
  {\bibfield  {journal} {\bibinfo  {journal} {Physical Review Letters}\
  }\textbf {\bibinfo {volume} {107}},\ \bibinfo {pages} {203602} (\bibinfo
  {year} {2011})}\BibitemShut {NoStop}%
\bibitem [{\citenamefont {Ruffner}\ and\ \citenamefont
  {Grier}(2012)}]{ruffner_optical_2012}%
  \BibitemOpen
  \bibfield  {author} {\bibinfo {author} {\bibfnamefont {D.~B.}\ \bibnamefont
  {Ruffner}}\ and\ \bibinfo {author} {\bibfnamefont {D.~G.}\ \bibnamefont
  {Grier}},\ }\href {https://doi.org/10.1103/PhysRevLett.109.163903} {\bibfield
   {journal} {\bibinfo  {journal} {Physical Review Letters}\ }\textbf {\bibinfo
  {volume} {109}},\ \bibinfo {pages} {163903} (\bibinfo {year}
  {2012})}\BibitemShut {NoStop}%
\bibitem [{\citenamefont {Boyd}\ and\ \citenamefont
  {Vandenberghe}(2004)}]{boyd_convex_2004}%
  \BibitemOpen
  \bibfield  {author} {\bibinfo {author} {\bibfnamefont {S.~P.}\ \bibnamefont
  {Boyd}}\ and\ \bibinfo {author} {\bibfnamefont {L.}~\bibnamefont
  {Vandenberghe}},\ }\href@noop {} {\emph {\bibinfo {title} {Convex
  optimization}}}\ (\bibinfo  {publisher} {Cambridge University Press},\
  \bibinfo {address} {Cambridge, UK ; New York},\ \bibinfo {year}
  {2004})\BibitemShut {NoStop}%
\end{thebibliography}

%

\end{document}